\documentstyle[titlepage,12pt]{article}
\begin{document}
\newcommand{\be}{\begin{equation}}
\newcommand{\ee}{\end{equation}}
\newcommand{\pbb}{\bar{\cal P}}
\newcommand{\pcc}{{\cal P}}
\title{The relativistic top: An application
of the BFV quantization procedure for systems with degenerate constraints}

\author{N. K. Nielsen\thanks{e-mail address: nkn@fysik.ou.dk}
and U. J. Quaade\thanks{e-mail address: ujq@dou.dk}, \\
Fysisk Institut, Odense University, Denmark.}

\begin{titlepage}
\maketitle

\begin{abstract}
The physical phase space of the relativistic top, as defined by Hanson and
Regge,
is expressed in terms of canonical coordinates of the  Poincar\'e group
manifold.
The system is described
in the Hamiltonian formalism by the mass shell condition and constraints that
reduce the number of spin degrees of freedom.

The constraints are second class and are modified into a set of first class
constraints by adding combinations of gauge fixing functions. The
Batalin-Fradkin-Vilkovisky (BFV) method
is then applied to quantize the system in the
path integral formalism in Hamiltonian form. It is finally shown that
different gauge choices produce different equivalent forms
of the constraints.
\end{abstract}
\end{titlepage}
\section{Introduction}
In a classical paper by Hansson and Regge \cite{hansreg}
(cf. also \cite{nieto})
a Hamiltonian
formulation of the relativistic spherical top was derived. The description
contains  too many degrees of freedom and constraints are imposed.
In the Hamiltonian formulation the system is then fully
described by the constraints for the spin, which can be imposed in
the form suggested by Pryce:
\be
S_{\mu\nu}P^\nu=0  \label{pryce}
\ee
as well as the mass shell condition that has the form
\cite{hansreg}\cite{nieto}:
\begin{equation}
P^2+\frac{1}{2}S^{\mu \nu }S_{\mu \nu }=0 \label{trac}
\end{equation}
($P^{\mu }$ and $S^{\mu \nu }$ are the momentum and spin variables,
respectively. Our metric is
 $$
g_{\mu \nu }={\rm diag}(-1, 1, \cdots ,1). $$
We  work throughout the paper in an arbitrary dimensionality
$D$ of space-time.).

In \cite{hansreg} the system was quantized by means of the Dirac method
which implies
an explicit realization of the physical subspace before quantization. This
means that gauge conditions have to be imposed and new physical variables
defined that obey canonical Poisson brackets. Manifest Lorentz
covariance is in general lost by this procedure.

More modern quantization approaches are the BRST\cite{BRST} or BFV\cite{bfv}
quantization procedures.
An attempt to use these methods on the top was carried out in \cite{malik}.
However, the non-commutative properties of the spin variables as well as the
degeneracy of their constraints  were ignored in
\cite{malik}. In the present paper, these complications of the problem are
fully considered.
We do this by using as coordinates the canonical coordinates of the
Poincar\'e group manifold. Sec.2 of the paper
is devoted to this construction.

The spin constraints (1) are second class constraints. We first deal with
this difficulty in the simpler case of a massive spinning particle with
the mass-shell constraint:
\begin{equation}
P^2+m^2=0
\end{equation}
($m$ is a constant). In this case, we convert the spin constraints to
first class constraints by replacing them with:
\begin{equation}
\psi _{\mu }=S_{\mu \nu}(P^{\nu }-m\Lambda ^{0\nu})=0,
\end{equation}
where $\Lambda ^{0\nu}$ is a Lorentz transformation matrix element.
Different gauge choices will correspond to the various versions of
the spin constraints as reviewed in \cite{hansreg} App. B.

The spin constraints are degenerate since
\be
(P^{\mu }-m\Lambda ^{0\mu})\psi _{\mu }=0.
\ee
In order to quantize the system in a manifestly covariant way we thus have
to use the version of
BFV quantization
appropriate for degenerate constraints \cite{deg}. The details are contained
in sec.3 of the paper.

In sec.4 we finally consider the quantization of the relativistic top.
The complete set of first class constaints is here determined as
\be
S_{\mu \nu}(P^{\nu }-\eta \Lambda ^{0\nu}),
\ee
\be
P^2+\eta ^2
\ee
and
\be
\eta ^2-\frac{1}{2}S^{\mu \nu }S_{\mu \nu }
-S^{\nu \lambda }\Lambda ^0\hspace{0.1 mm}_{\lambda }
S_{\nu \rho }\Lambda ^{0\rho }.
\ee
We have found it convenient to introduce an extra auxiliary variable and an
extra constraint with conjugate momentum $\eta $.
We then demonstrate how various gauge choices
lead to the appriate forms of the mass shell constraint (\ref{trac}).

An appendix contains supplementary material to sec.2.

Throughout the paper we work with Poisson brackets. The quantum mechanical
commutators or anticommutators are then
obtained by the usual replacement.

\section{Group Theoretical Setting}

For an arbitrary Lie group\footnote{For a summary of the basic concepts see
for example  \cite{loch}.} with structure constants
$C^{\gamma }\hspace{0.1 mm}_{\alpha \beta }$
we introduce the matrix $\phi _{\alpha }\hspace{0.1 mm}^{\gamma }$:
\begin{equation}
\phi _{\alpha }\hspace{0.1 mm}^{\gamma }
=C^{\gamma }\hspace{0.1 mm}_{\beta \alpha }\xi^{\beta } \label{phi}
\end{equation}
where $\xi^{\beta }$ is a canonical coordinate on the group manifold.
The inverse vielbein $u_{\alpha }\hspace{0.1 mm}^{\beta }$ will then
near the origin of the group manifold be given by:
\begin{equation}
u_{\alpha }\hspace{0.1 mm}^{\beta }
=(\frac{e^{\phi }-1}{\phi })_{\alpha }\hspace{0.1 mm}^{\beta }
\end{equation}
and hence fulfil the Cartan-Maurer equation:
\begin{equation}
\partial _{\alpha }u_{\beta }\hspace{0.1 mm}^{\gamma }-
\partial _{\beta }u_{\alpha }\hspace{0.1 mm}^{\gamma }=
u_{\alpha }\hspace{0.1 mm}^{\epsilon }u_{\beta }
\hspace{0.1 mm}^{\delta }C^{\gamma
}\hspace{0.1 mm}_{\epsilon \delta}.
\end{equation}
The conjugate momentum to the group manifold coordinate
$\xi ^{\beta }$ is $\Pi _{\alpha }$ and is related to the generator
$I_{\beta }$ through:
\begin{equation}
\Pi _{\alpha }=u_{\alpha }\hspace{0.1 mm}^{\beta }I_{\beta }.
\end{equation}
Poisson brackets are found from:
\begin{equation}
\{\Pi_{\alpha },\xi ^{\beta}\}=-\delta
_{\alpha }\hspace{0.1 mm}^{\beta }.  \label {Ipi}
\end{equation}
The generators fulfil, by the Cartan-Maurer
equation:
\begin{eqnarray}
\{I_{\alpha } , I_{\beta }\}
=C^{\gamma }\hspace{0.1 mm}_{\alpha \beta }I _{\gamma }.
\end{eqnarray}

In \cite{hansreg} the description of the relativistic top was made in terms of
physical quantities such as the spacetime coordinate $x^\mu$ and the
conjugate momentum $P_\mu$ together with Lorentz transformations
$\Lambda^{\mu }\hspace{0.1 mm}_{\nu}$ and the spin variables $S_{\mu\nu}$.
The description can however be
made on a more fundamental level in terms of the
canonical coordinates of the
Poincar\'e group manifold and their conjugate momenta.

The Poincar\'e group generators are denoted $P_{\mu }$ (translations)
and $M_{\mu \nu }$
(rotations and proper Lorentz transformations), and the nonvanishing structure
constants are:
\begin{equation}
C^{\xi }\hspace{0.1 mm}_{\mu \nu ,\sigma}=\delta ^{\xi
}\hspace{0.1 mm}_{\nu }\eta _{\mu \sigma}-\delta ^{\xi }\hspace{0.1 mm}_{\mu
}\eta _{\nu
\sigma }
\end{equation}
and
\begin{equation}
C^{\xi \eta }\hspace{0.1 mm}_{\mu \nu, \sigma \rho}=
\delta ^{\xi }\hspace{0.1 mm}_{\sigma }
C^{\eta }\hspace{0.1 mm}_{\mu \nu ,\rho }
-\delta ^{\eta }\hspace{0.1 mm}_{\sigma }
C^{\xi }\hspace{0.1 mm}_{\mu \nu ,\rho }
-\delta ^{\xi }\hspace{0.1 mm}_{\rho }
C^{\eta }\hspace{0.1 mm}_{\mu \nu ,\sigma }
+\delta ^{\eta }\hspace{0.1 mm}_{\rho }
C^{\xi }\hspace{0.1 mm}_{\mu \nu ,\sigma }.
\end{equation}
The Poincare group is parametrized by the variables
$a^{\mu }$ (translations) and $\lambda ^{\mu \nu }=-\lambda ^{\nu \mu }$
(rotations and proper Lorentz transformations).
The corresponding conjugate momenta are denoted $\Pi _{\mu }$ and
$\Pi _{\mu \nu }$, respectively.
With:
\begin{equation}
\phi _{\lambda }\hspace{0.1 mm}^{\rho }=\frac{1}{2}
\lambda^{\mu \nu }C^{\rho }\hspace{0.1 mm}_{\mu \nu , \lambda }. \label{fi}
\end{equation}
a proper Lorentz transformation is given by:
\begin{equation}
\Lambda ^{\lambda }\hspace{0.1 mm}_{\sigma }
=(e^{-\phi })_{\sigma }\hspace{0.1 mm}^{\lambda }.  \label{lorentz}
\end{equation}
The inverse vielbeins $u_{\sigma \rho }\hspace{0.1 mm}^{\lambda \tau },
u_{\sigma }\hspace{0.1 mm}^{\lambda
}$ and $u_{\mu
\nu}\hspace{0.1 mm}^{\lambda}$ are constructed from the structure
constants as described above. The generators and momenta are related through
(cf. (\ref{Ipi}))
\begin{equation}
\Pi _{\mu }=u_{\alpha }\hspace{0.1 mm}^{\nu }P_{\nu }, \label{vielI}
\end{equation}
\begin{equation}
\Pi _{\mu \nu }
=\frac{1}{2}u_{\mu \nu }\hspace{0.1 mm}^{\lambda \rho }M_{\lambda \rho }
+u_{\mu \nu }\hspace{0.1 mm}^{\lambda }P_{\lambda }. \label{vielII}
\end{equation}
This ensures by the Cartan-Maurer equation
that the generators fulfil
the appropriate Poisson brackets relations:
\begin{equation}
\{M_{\mu \nu}, M_{\sigma  \rho}\}
=\frac{1}{2}C^{\xi \eta }\hspace{0.1 mm}_{\mu \nu, \sigma \rho}
M_{\xi \eta }, \label{M, M}
\end{equation}
\begin{equation}
\{M_{\mu \nu}, P_{\sigma  }\}=C^{\xi
}\hspace{0.1 mm}_{\mu \nu ,\sigma }P_{\xi },
\end{equation}
\begin{equation}
\{P_{\mu }, P_{\nu }\}=0. \label{P, P}
\end{equation}
The physical space-time
coordinates, denoted $x ^{\lambda }$, have to be defined through:
\begin{equation}
x ^{\lambda }=a^{\nu }u_{\nu }\hspace{0.1 mm}^{\lambda } \label{vielIII}
\end{equation}
in order to obey the Poisson bracket relations:
\begin{equation}
\{x ^{\mu }, P _{\nu }\}=\delta _{\nu }\hspace{0.1 mm}^{\mu}. \label {x, P}
\end{equation}

The spin variables $S_{\mu \nu }$ are introduced through the following
decomposition of $M_{\mu \nu }$:
\begin{equation}
M_{\mu \nu }=L_{\mu \nu }+S_{\mu \nu }
\end{equation}
with
\begin{equation}
L_{\mu \nu }=x _{\mu }P_{\nu }-x _{\nu }P_{\mu },
\end{equation}
Poisson brackets involving $L_{\mu \nu}$ are:
\begin{equation}
\{L_{\mu \nu}, P_{\sigma  }\}=C^{\lambda }\hspace{0.1 mm}_{\mu \nu ,\sigma
}P_{\lambda  },
\end{equation}
\begin{equation}
\{L_{\mu \nu}, x_{\sigma  }\}
=C^{\lambda }\hspace{0.1 mm}_{\mu \nu ,\sigma }x _{\lambda  },
\end{equation}
and
\begin{eqnarray}
\{L_{\mu \nu}, L_{\sigma  \rho}\}=\frac{1}{2}C^{\xi \eta }\hspace{0.1
mm}_{\mu \nu, \sigma
\rho}L_{\xi \eta },
\end{eqnarray}
Therefore:
\begin{equation}
\{S_{\mu \nu}, P_{\sigma  }\}=0. \label{S, P}
\end{equation}
In the appendix we also prove:
\begin{equation}
\{S_{\mu \nu}, x _{\sigma  }\}=0.  \label {S, x}
\end{equation}
By combination of these Poisson bracket relations we get:
\begin{eqnarray}
\{S_{\mu \nu}, S_{\sigma  \rho}\}
=\frac{1}{2}C^{\xi \eta }\hspace{0.1 mm}_{\mu \nu, \sigma
\rho}S_{\xi \eta }. \label{S, S}
\end{eqnarray}
For a Lorentz transformation
$\Lambda ^{\tau }\hspace{0.1 mm}_{\sigma  }$
we also prove in the appendix:
\begin{eqnarray}
\{S_{\mu \nu}, \Lambda ^{\tau }\hspace{0.1 mm}_{\sigma  }\}
=C^{\zeta }\hspace{0.1 mm}_{\mu \nu ,\sigma }\Lambda ^{\tau }\hspace{0.1
mm}_{\zeta  }. \label {S, Lambda}
\end{eqnarray}
Finally we list the vanishing Poisson brackets:
\be
\{x_{\mu  }, x_{\nu  }\}=\{\Lambda ^{\mu }\hspace{0.1 mm}_{\nu  },
\Lambda ^{\tau }\hspace{0.1 mm}_{\sigma  }\}
= \{x_{\mu  },\Lambda ^{\tau }\hspace{0.1 mm}_{\sigma  }\}=\{P_{\mu
},\Lambda ^{\tau }\hspace{0.1 mm}_{\sigma  }\}=0. \label{nul}
\ee

We also introduce the associated quantity
$\sigma ^{\mu \nu }$ :
\begin{equation}
\sigma ^{\mu \nu}=(\Lambda ^{-1})^{\mu
}\hspace{0.1 mm}_{\lambda }\dot{\Lambda }^{\lambda \nu
}=\frac{1}{2}\dot{\lambda }^{\rho \sigma
}u_{\rho \sigma }\hspace{0.1 mm}^{\mu \nu }.
\end{equation}
which is the space-time equivalent to the angular velocity in Euclidean space.
The classical action $S$ in first order formalism is:
\be
S=\int d\tau (\dot{a}^{\alpha }\Pi _{\alpha }+\frac{1}{2}\dot{\lambda
}^{\mu \nu }
\Pi _{\mu \nu })        \label{action}
\ee
where from (\ref{vielI}) and (\ref{vielII})
\be
\dot{a}^{\alpha }\Pi _{\alpha }+\frac{1}{2}\dot{\lambda }^{\mu \nu }
\Pi _{\mu \nu }
=\dot{x }^{\alpha }P_{\alpha }+\frac{1}{2}\sigma ^{\mu \nu }S_{\mu \nu }
\ee
The action (\ref{action}) describes a particle moving on the Poincar\'e
group manifold parametrized by canonical coordinates.

The main point of this section is that we have demonstrated the convenience
of the canonical Poincar\'{e} group coordinates and their conjugate momenta
for the description
of the spinning top. Our approach is equivalent to that of Hanson and Regge
\cite{hansreg} (thus our Poisson bracket relations (\ref{P, P}), (\ref{x, P}),
(\ref{S, P}), (\ref{S, x}),
(\ref{S, S}), (\ref{S, Lambda}) and (\ref{nul}) are identical
to \cite{hansreg} eq.
(3.11)). However, by avoiding the use of
$\Lambda ^{\mu }\hspace{0.1 mm}_{\nu }$ as a canonical coordinate we also
avoid the difficulties related to the
constraint $\Lambda \Lambda ^T=g$.

We conclude this section by listing the Poisson brackets that are required
in sec.3 and 4:
\begin{eqnarray}
\{S_{\mu \nu}, S_{\sigma  \rho}\}=\eta _{\mu \sigma }S_{\nu \rho }
-\eta _{\nu \sigma }S_{\mu \rho }
+\eta _{\nu \rho }S_{\mu \sigma }-\eta _{\mu \rho }S_{\nu \sigma },
\nonumber \\
\{S_{\mu \nu}, \Lambda ^{\tau }\hspace{0.1 mm}_{\sigma  }\}
=\eta _{\mu \sigma }\Lambda ^{\tau }\hspace{0.1 mm}_{\nu  }
-\eta _{\nu \sigma }\Lambda ^{\tau }\hspace{0.1 mm}_{\mu  },
\nonumber \\
\{P_{\mu }, P_{\nu }\}=\{S_{\mu \nu}, P_{\sigma  }\}=\{P_{\mu
},\Lambda ^{\tau }\hspace{0.1 mm}_{\sigma  }\}
=\{\Lambda ^{\mu }\hspace{0.1 mm}_{\nu  },
\Lambda ^{\tau }\hspace{0.1 mm}_{\sigma  }\}=0. \label {collect}
\end{eqnarray}

\section{Quantization of massive spinning particle}
In \cite{hansreg} the relativistic top was shown to be fully described in
terms a set of constraints for the spin variables
\setcounter{equation}{0}
\be
\raggedleft S_{\mu\nu}P^\nu=0,
\ee
and the mass shell constraint
\be
P^2+\frac{1}{2}S^{\mu \nu }S_{\mu \nu }=0.
\ee
\setcounter{equation}{39}
When quantizing the system by the BFV method there are two difficulties.
First the constraints are not linearly independent since
\be
P^\mu\psi_\mu=0.
\ee

In the framework of \cite{deg}
the system is called first stage reducible. The system can be quantized using
the method of \cite{deg}, provided the constraints are first class.
This brings us to the second difficulty. Inspecting the mutual
Poisson brackets of $\psi_\mu$ by means of (\ref{collect}) we find
\be
\{\psi_\mu,\psi_\nu\}
=P^2S_{\mu\nu}+P_\mu\psi_\nu-P_\nu\psi_\mu \label{algebra},
\ee
and we see from the first term on the right-hand side that the constraints
are second class.

To see how this obstacle can be overcome, we consider in this section
the simpler case of a massive spinning particle with the  usual mass-shell
constraint:
\be
P^2+m^2=0
\ee
with $m$ a fixed mass parameter, postponing the treatment of the
constraint (\ref{trac}) to sec.4. From (\ref {algebra}) we see that
the constraints are now first class for $m=0$. To make the constraints
first class also for $m\neq 0$ we modify them as follows:
\be
\psi_{\mu }=S_{\mu \nu}(P^{\nu }-m\Lambda ^{0\nu}).  \label{modpryce}
\ee
The Poisson brackets of the new constraints are again obtained
from (\ref{collect}) supplemented with $\Lambda \Lambda ^T=g$:
\be
\{\psi_{\mu }, \psi_{\nu }\}=(P^2+m^2)S_{\mu \nu }+
P_{\mu }\psi_{\nu }-P_{\nu }\psi_{\mu }.
\ee
The constraints are now first class. Notice that
the new constraints are also manifestly Lorentz invariant since $\Lambda $
transforms as a Lorentz vector under $S_{\mu \nu }$ only in its second index
(cf. (\ref {S, Lambda}), and \cite{hansreg}, remark after eq. (3.11)).

The system is still first stage reducible and the reducibility condition
now reads
\be
(P^{\mu }-m\Lambda ^{0\mu})\psi_\mu=0
\ee

Quantization of the system is carried out using the BFV method for the case of
degenerate constraints \cite{deg}.
Ghosts and Lagrange multipliers and their corresponding conjugate momenta
are introduced. In the following, all the new degrees of freedom are listed,
with coordinates first and conjugate momenta last in $(\cdot ,\cdot )$.

Ghosts and antighosts corresponding to the constraints $\psi _{\mu}$ are
(ghostnumber 1,-1 respectively)
\be
(c^\mu,\pbb_\nu); \ \ \ (\pcc^\nu , {\bar c}_\mu ),
\ee
and to the mass-shell constraint
\be
(c,\pbb); \ \ \  (\pcc , {\bar c}).
\ee
All these ghosts and antighosts obey Fermi statistics.
Lagrange multipliers (that are bosonic) for the constraints and their
conjugate momenta  are introduced
\be
(\lambda^\mu,\pi_\nu);\ \ \ \ (e,\pi). \label{g3}
\ee

Ghosts and antighosts corresponding to reducibility of the spin
constraints (ghostnumber 2,-2) are
\be
(c^\prime,\pbb^\prime);  \ \ \ (\pcc^\prime , {\bar c}^\prime).  \label{g2}
\ee
These ghosts and antighosts obey Bose statistics. For the ghost variables of
the spin constraints we also need a constraint and hence a Lagrange multiplier
and its conjugate momentum (that obey Fermi statistics):
\be
(\lambda^\prime,\pi^\prime).
\ee

An extra Lagrange multiplier and its conjugate momentum
\be
(\lambda^{\prime\prime},\pi^{\prime\prime})
\ee
(with Bose statistics) are furthermore introduced
to fix the gauges of ($\lambda^\mu , \pi _\nu $).
Likewise, an extra ghost and its conjugate momentum
\be
({\bar c}^{\prime\prime},\pcc^{\prime\prime})   \label{g4}
\ee
(with Fermi statistics) fix the gauges of  ($\pcc^\nu, {\bar c}_\mu$).

The complete BRST charge $Q$ is now determined by
the procedure described in \cite{deg}
\be
Q=Q_{min}+\pi_\mu\pcc^\mu+\pi\pcc+\pi^\prime\pcc^\prime
+\pi^{\prime\prime}\pcc^{\prime\prime}
\ee
with
$$
Q_{min}=\psi_\mu c^\mu+(P^2+m^2)c
$$
$$
-P_\mu c^\mu c^\nu\pbb_\nu+(P^\nu-m\Lambda^{0\nu})c^\prime\pbb_\nu
-\frac12S_{\mu\nu}c^\mu c^\nu\pbb
+(P_\mu+m\Lambda^0_{\ \mu})c^\mu c^\prime\pbb^\prime
$$
\be
+c^\mu c^\prime\pbb_\mu\pbb-(c^\prime)^2\pbb\pbb^\prime.
\ee
It is nilpotent:
\be
\{Q, Q\}=0
\ee
because the constraints are first class.

The presence of structure functions (not constants) in the
constraint algebra will in general give rise to higher order ghost
terms in the BRST
charge. However, in the present instance all the higher order structure
functions turn out to vanish.

$Q$ has nonvanishing Poisson brackets with the generators
$M_{\mu \nu }$ of rotations and proper Lorentz transformations.
However, we can redefine $M_{\mu \nu }$ by addition of
terms involving ghost variables:
\be
M_{\mu \nu }\rightarrow M_{\mu \nu }-c_{\mu }\pbb_{\nu }+c_{\nu }\pbb_{\mu }
\ee
such that the redefined generators have vanishing Poisson brackets
with $Q$ and still fulfil the appropriate structure relations (\ref {M, M}).
This shows that also the quantized theory is Lorentz invariant.

The Hamiltonian $H$ of the theory is given by the general
expression \cite{deg}:
\be
H=\{\Psi, Q\}
\ee
where $\Psi $ is the gauge fermion (since the theory is determined by
its constraints \cite{hansreg},
no further terms occur in $H$). $H$ 	has vanishing
Poisson bracket with the BRST charge $Q$:
\be
\{H, Q\}=0
\ee
because of the nilpotency of $Q$. The general form of $\Psi$ is:
\be
\Psi={\bar c}_\mu(\chi^\mu+{\dot\lambda}^\mu)
+{\bar c}(\chi+{\dot e})+{\bar c}^\prime(\chi^\prime +{\dot\lambda}^{\prime })
+{\bar c}^{\prime \prime }(\chi^{\prime \prime }
+{\dot\lambda}^{\prime \prime })
+\lambda^{\prime \prime }G
+\pbb_\mu\lambda^\mu+\pbb e+\pbb^\prime\lambda^\prime.
\label{gf}
\ee
Here $\chi^\mu $ and $\chi $ are the gauge fixing functions corresponding to
the spin constraints and the mass shell condition, respectively. $\chi^\prime$
fixes the  gauge for the ghosts $c^\mu$ that are gauge variables due
to the reducibility of the spin constraints.
$\chi^{\prime \prime}$ can be considered a gauge fixing function for the
Lagrange multipliers $\lambda^\mu$ and
$G$ a gauge fixing function for the antighosts ${\bar c}_\mu$.

With this choice of $\Psi $  the Hamiltonian $H$ has the form
$$
H=H_1+H_2+H_3+H_4,
$$
$$
H_1=e(P^2+m^2)+\lambda ^{\mu }(\psi _{\mu }-S_{\mu \nu }c^{\nu }\pbb
-c^{\nu }(\pbb_{\nu }P_{\mu }-\pbb_{\mu }P_{\nu })
-c^{\prime }\pbb\pbb_{\mu }
+c^{\prime }\pbb^{\prime } (P_{\mu }-m\Lambda^0_{\ \mu}))
$$
$$
-\lambda ^{\prime }((P^{\mu }-m\Lambda ^{0\mu })\pbb_{\mu }
+\pbb c^{\mu }\pbb _{\mu }+\pbb^{\prime }c^{\mu }(P_{\mu }
+m\Lambda ^0\hspace{0.1 mm}_{\mu })
-2c^{\prime }\pbb^{\prime }\pbb),
$$
$$
H_2=\pbb \pcc+\pbb _\mu \pcc ^\mu +\pbb ^{\prime }\pcc ^{\prime },
$$
$$
H_3=\pi \dot{e}+\pi _{\mu }\dot{\lambda }^{\mu }
+\pi ^{\prime }\dot{\lambda }^{\prime }
+\pi ^{\prime \prime }\dot{\lambda }^{\prime \prime }
+\pcc \dot{\bar{c}}+\pcc ^{\mu }\dot{\bar{c}}_{\mu }
+\pcc ^{\prime }\dot{\bar{c}}^{\prime }
+\pcc ^{\prime \prime }\dot{\bar{c}}^{\prime \prime }
$$
$$
+\frac{d}{d\tau }(\bar{c}\pcc+\bar{c}_{\mu }\pcc ^{\mu }
+\bar{c}^{\prime }\pcc ^{\prime }
+\bar{c}^{\prime \prime }\pcc ^{\prime \prime }),
$$
\be
H_4=\chi \pi +\chi ^{\mu }\pi _{\mu }
+\chi ^{\prime }\pi ^{\prime }+\chi ^{\prime \prime}\pi ^{\prime \prime }
+G\pcc ^{\prime \prime }
+\bar{c}\{\chi , Q\}+\bar{c}_{\mu }\{\chi ^{\mu }, Q\}
+\bar{c}^{\prime }\{\chi ^{\prime }, Q\}
+\bar{c}^{\prime \prime}\{\chi ^{\prime \prime}, Q\}
+\lambda^{\prime \prime}\{G, Q\}.
\ee

This expression is quite complicated; the important thing about it is
that $H_1$
contains a linear combination of the constraints, while $H_4$ contains a
linear combination of the gauge fixing functions.

The familiar possibilities for choosing the parameter $\tau $ emerge
through the gauge choices:
\be
\chi=x^0-\tau \hspace{1 cm}{\rm(coordinate\ time\ gauge)}, \label{tid}
\ee
\be
\chi=x^+-\tau \hspace{1 cm}{\rm (light-cone\ gauge)}
\ee
and
\be
\chi=\dot{x}^2+1 \hspace{1 cm}{\rm (covariant\ proper\ time\ gauge)}.
\ee

The gauge choices corresponding to the spin constraints are  more interesting.
In \cite{hansreg} three equivalent forms of the spin constraints
were mentioned. One is:
\setcounter{equation}{0}
\be
\raggedleft S_{\mu\nu}P^\nu=0,
\ee
\setcounter{equation}{63}
(often referred to as the Pryce constraint). It emerges from (\ref{modpryce})
by the gauge choice (cf. \cite{hansreg}):
\be
\chi_\mu=\Lambda^0_{\ \mu}+\frac{P_\mu}{m}=0.
\ee
Another possible version of the spin constraints is:
\be
S^{0j}-\frac{P^l}{P^0+m}S^{lj}=0,
\ee
that arises naturally in Wigner's classification of the representations of the
Poincar\'{e} group \cite{wigner}. It can be obtained from (\ref{modpryce})
by the gauge choice
\be
\chi_\mu=
\Lambda^0_{\ \mu}-\delta^0\hspace{0.1 mm}_\mu=0. \label{wigner}
\ee
Finally, the spin constraints
\be
 S^{0\mu}=0
\ee
result from (\ref{modpryce}) by the gauge choice
\be
\chi_\mu =\Lambda^0_{\ \mu}-\delta^0\hspace{0.1 mm}_\mu-\frac{P_\mu}{m}=0.
\ee
This means that different forms of the constraint (\ref{modpryce}) are
obtained
for different gauge choices. This nice feature comes from the fact that we
were forced to make the constraints gauge dependent through the modification
(\ref{modpryce}) in order to make the constraint algebra first class.

In the case $m=0$ it is customary to use light-cone quantization which
emerges by the gauge choice
\be
\chi_\mu=\Lambda^+_{\ \mu}-\delta^+\hspace{0.1 mm}_\mu=0.
\ee

The particle propagator is given by
\be
\langle x, \lambda, t \mid x^{\prime }, \lambda^{\prime }, 0\rangle =
\int [d^nP][d^nQ]\exp
\{i\int_0^t d\tau(P_A{\dot Q}^A-H)\} \label{path}
\ee
where $dP$ and $dQ$ denotes integration in all variables: physical, ghosts
and Lagrange multipliers. Likewise the sum $P_A{\dot Q}^A$ runs over all
variables.
The Hamiltonian $H$ provides the gauge fixing, and it is known \cite{deg}
that  $\langle x, \lambda, \tau \mid x^{\prime },
\lambda^{\prime }, 0\rangle $ is independent of the
choice of the gauge fermion $\Psi$.

We conclude this section by computing the propagator from the above
expression for the gauge conditions (\ref{tid}) and (\ref{wigner})
supplemented with $\chi ^{\prime }=c^0$, $\chi ^{\prime \prime }=\lambda ^0$
and $G=\pbb ^0$. As was mentioned above, this gauge choice is preferred
in connection with Wigner's analysis of the representations of the
Poincar\'{e} group. After integration over ghost and Lagrange multiplier
variables and their conjugate momenta we obtain the expression:
\be
\langle x, \lambda, t \mid x^{\prime }, \lambda^{\prime }, 0\rangle =
\int  [d^Da][d^D\Pi][d^{\frac{D(D-1)}{2}}\lambda]
[d^{\frac{D(D-1)}{2}}\Pi](\prod A)e^{i\int _0^t d\tau({\dot x}^\mu P_\mu
+\frac12\sigma^{\mu\nu}S_{\mu\nu})}
\ee
with
\be
A=2P^0 \delta (x^0-\tau)\delta (P^2+m^2)\prod _{j=1}^{D-1}
\delta (\Lambda ^0\hspace{0.1  mm}_j)\delta (S^{0j}-\frac{P^l}{P^0+m}S^{lj}).
\ee
Here and in (\ref{path}) the time interval $[0, t]$ has been cut in small
subintervals
in the standard manner, with one factor $A$ for each subinterval.

The integrations over the coordinates $\lambda ^{\mu \nu }$
and their conjugate momenta are eliminated by means of (\ref{lorentz}) and
(\ref {vielII}) and by using in the initial
and final states of the path integral instead of configurations of
$\lambda $-coordinates eigenstates with eigenvalues
labelled $\{n\}$ and $\{n^{\prime }\}$ of the Cartan subalgebra
of the little group Lie algebra in a particular representation.
Using finally (\ref{vielI}) and (\ref{vielII}) to
change integration variables variables we obtain the standard expression
for the propagator (cf. \cite{marnelius}):
\be
\langle x, \{n\}, t \mid x^{\prime }, \{n\}, 0\rangle =
\delta _{\{n\}, \{n^{\prime }\}}
\int [d^Dx][d^DP](\prod 2P^0\delta (x^0-\tau)\delta (P^2+m^2))
e^{i\int_0^t d\tau{\dot x}^\mu P_\mu }
\ee
with $\delta _{\{n\}, \{n^{\prime }\}}$ Kronecker's $\delta $.

\section{The relativistic top}

We now turn to our main concern: BFV quantization of the relativistic top.

The relativistic top is, as noted in the introduction, fully described by
the two constraints  (\ref{pryce}) and (\ref{trac}). The constraint algebra
does not close,
and the modification (\ref{modpryce}) alone is not sufficient to make the
constraint algebra close. To see how this comes about, we compute
the Poisson bracket:
\be
\{\frac{1}{2}S^{\lambda \nu }S_{\lambda \nu }, S_{\mu \rho }\Lambda ^{0\rho }\}
=2S_{\mu \lambda }S^{\lambda \rho }\Lambda ^0\hspace{0.1 mm}_{\rho },
\ee
where again (\ref {collect}) was used.
This result shows that the modified spin constraints
(\ref{modpryce}) combined with (\ref{trac}) do not constitute a set of first
class constraints.

However, (\ref {collect}) also imply the Poisson bracket relation:
\be
\{S^{\nu \lambda }\Lambda ^0\hspace{0.1 mm}_{\lambda }
S_{\nu \rho }\Lambda ^{0 \rho },
S_{\mu \sigma }\Lambda ^{0\sigma }\}
=-2S_{\mu \lambda }S^{\lambda \rho }\Lambda ^0\hspace{0.1 mm}_{\rho }.
\ee
Thus, the combination
\be
\frac{1}{2}S^{\mu \nu }S_{\mu \nu }+S^{\nu \lambda }
\Lambda ^0\hspace{0.1 mm}_{\lambda }
S_{\nu \rho }\Lambda ^{0 \rho } \label{sum}
\ee
has vanishing Poisson bracket with
$
S_{\mu \sigma }\Lambda ^{0\sigma }.
$
Furthermore it follows from (\ref{collect}) that
\be
\{S^{\nu \lambda }\Lambda ^0\hspace{0.1 mm}_{\lambda }
S_{\nu \rho }\Lambda ^{0 \rho },
S_{\mu \sigma }\}=0
\ee
i.e. the quantity $S^{\nu \lambda }\Lambda ^0\hspace{0.1 mm}_{\lambda }
S_{\nu \rho }\Lambda ^{0 \rho }$ is a Lorentz scalar, as
is also, by the same indication,  $\frac{1}{2}S^{\mu \nu }S_{\mu \nu }$.

A possible mass shell constraint is thus
\be
P^2+\frac{1}{2}S^{\mu \nu }S_{\mu \nu }+S^{\nu \lambda }
\Lambda ^0\hspace{0.1 mm}_{\lambda }
S_{\nu \rho }\Lambda ^{0 \rho }=0 \label{comb}
\ee
Before we turn to the final formulation of the constraints, we discuss
the appearance of (\ref{comb}) in the gauges listed in sec.3.

For $\Lambda ^{0}\hspace{0.1 mm}_{\mu }=-\frac{P_{\mu }}{m}$ the spin
constraints reduce to $S_{\mu \nu }P^{\nu }=0$; this means that
(\ref{comb}) reduces to (\ref{trac}).

In the gauge where $\Lambda ^{0}\hspace{0.1 mm}_{\mu }
=\delta^0\hspace{0.1 mm}_\mu $
we get
\be
\frac{1}{2}S^{\mu \nu }S_{\mu \nu }
+S^{\nu \lambda }\Lambda ^0\hspace{0.1 mm}_{\lambda }
S_{\nu \rho }\Lambda ^{0 \rho }=\frac{1}{2}S^{ij}S_{ij},
\ee
where the summation on the right hand side runs over only spatial indices, i.e.
the right hand side is the square of the spin in the ordinary spatial
sense.
The mass shell constraint (\ref{comb}) thus becomes
\be
P^2+\frac{1}{2}S^{ij}S_{ij}=0 \label{regge}
\ee
As mentioned in sec.3, this gauge leads to Wigner's representation
theory of the Poincar\'{e} group, and the mass shell condition (\ref{regge})
gives when $\frac{1}{2}S^{ij}S_{ij}$ is replaced by its eigenvalue
a relationship between
mass and  spin quantum number. The mass shell condition (\ref{regge}) in
this gauge is equivalent to the condition (\ref{trac}) in the gauge where
$\Lambda ^{0}\hspace{0.1 mm}_{\mu }=-\frac{P_{\mu }}{m}$ (cf. \cite{hansreg},
remark after (3.60)).

We conclude that (\ref{comb}) is the proper replacement of (\ref{trac})
as the mass shell constraint for the relativistic top.

In order to generalize the constraints formulated in sec.3
to the top we introduce an auxiliary coordinate $\xi $ with the conjugate
momentum
variable $\eta $, and we introduce an extra constraint to eliminate again
the new degree of freedom. In this way we can avoid taking the square root of
(\ref{sum}). The new spin constraints are obtained from (\ref{modpryce})
by replacement of $m$ with $\eta $. The constraints of the relativistic
top are in this formulation:
\be
\psi_{\mu }=S_{\mu \nu}(P^{\nu }-\eta \Lambda ^{0\nu}),  \label{modmodpryce}
\ee
\be
{\cal M}=P^2 +\eta ^2,
\ee
and
\be
{\cal H}=\eta ^2-\frac{1}{2}S^{\mu \nu }S_{\mu \nu }
-S^{\nu \lambda }\Lambda ^0\hspace{0.1 mm}_{\lambda }
S_{\nu \rho }\Lambda ^{0 \rho }. \label{mass}
\ee
The mass shell constraint is ${\cal M}-{\cal H}$.

These constraints are first class since the Poisson brackets of ${\cal M}$ and
${\cal H}$ with all other constraints vanish, whereas
\be
\{\psi_{\mu }, \psi_{\nu }\}={\cal M}S_{\mu \nu }+
P_{\mu }\psi_{\nu }-P_{\nu }\psi_{\mu }.
\ee
The BFV quantization of the relativistic top is a trivial extension of the
BFV quantization of the massive spinning paticle in sec.3 because the new
constraint
${\cal H}$ has vanishing Poisson brackets with all other constraints.
Thus, BFV quantization is carried out  as in sec.3. The only new feature is
the extra variables corresponding to the constraint ${\cal H}$:
ghost and antighost variables
\be
(\tilde{c},\bar {\tilde{\cal P}}); \ \ \ (\tilde{\cal P},\bar {\tilde{c}}).
\ee
(that obey Fermi statistics), and a
Lagrange multiplier and its conjugate momentum
\be
(\tilde{\lambda },\tilde{\pi }).
\ee
The new BRST charge is
\be
Q=Q_{min}+\pi_\mu\pcc^\mu
+\pi\pcc+\tilde {\pi}\tilde{\pcc }+\pi^\prime\pcc^\prime
+\pi^{\prime\prime}\pcc^{\prime\prime}
\ee
with
$$
Q_{min}=\psi_\mu c^\mu+{\cal M}c+{\cal H}\tilde {c}
$$
$$
-P_\mu c^\mu c^\nu\pbb_\nu+(P^\nu-\eta \Lambda^{0\nu})c^\prime\pbb_\nu
-\frac12S_{\mu\nu}c^\mu c^\nu\pbb
+(P_\mu+\eta \Lambda^0_{\ \mu})c^\mu c^\prime\pbb^\prime
$$
\be
+c^\mu c^\prime\pbb_\mu\pbb-(c^\prime)^2\pbb\pbb^\prime.
\ee
Also, in the gauge fermion $\Psi $ we need now some extra terms:
\be
\Psi \rightarrow \Psi +{\bar {\tilde {c}}}(\tilde{\chi }
+\dot{\tilde {\lambda }})
+{\bar {\tilde {\cal P}}}\tilde{\lambda },
\ee
where $\tilde{\chi }$ is a new gauge fixing function that is coveniently
chosen according to:
\be
\tilde{\chi }=\xi
\ee
($\xi $ is the coordinate variable conjugate to $\eta $.).
The construction of the Hamiltonian and the propagator now runs exactly as in
sec.3; we
leave out the details.

It should be noticed that (\ref{mass}) can be generalized to
\be
{\cal H}=\eta ^2-f(\frac{1}{2}S^{\mu \nu }S_{\mu \nu }+S^{\nu \lambda }
\Lambda ^0\hspace{0.1 mm}_{\lambda }
S_{\nu \rho }\Lambda ^{0 \rho })
\ee
with $f$ an arbitrary function, without changing any of the
steps in the BFV construction. The corresponding general form of the
mass shell condition was the one considered in \cite{hansreg}.

\section{Conclusion}

Our main results in this paper are the following ones:

First, we have shown how the Poisson algebra relations (or commutators)
necessary
for the quantum theory of a spinning particle or a relativistic top
naturally emerge from the phase space related to the Poincar\'{e} group
manifold.

Secondly we have shown that the constraints defining a spinning particle or
a relativistic top, though initially second class, can be modified to make
them first class. What we use is that we are allowed to add to the
constraints combinations of gauge fixing functions. Accordingly, we show how
our modified constraints reduce to the previously known versions by specific
gauge choices.

Finally, since the systems we consider are entirely defined by their
(degenerate)
constraints, we have found it convenient to employ the BFV quantization
procedure
in Hamiltonian form, generalized to deal with degenerate constraints.
Our results give a simple and
yet nontrivial example of the application of this quantization scheme.

\vspace{2 mm}

{\large {\bf Acknowledgements:}} We thank J. Fredsted for discussions at the
early stages of this work. One of us (NKN) is grateful to E. Eilertsen for
teaching him the representations of the Poincar\'e group many years ago.
One of us (UJQ) is grateful to
R. Marnelius and B.-S. Skagerstam for helpful conversations, and the
Institute of Theoretical Physics,
Chalmers University of Technology, G\"{o}teborg, Sweden, where part of this
work was carried out, for hospitality.

\vspace{0.5 cm}

{\Large\bf{Appendix}}\\ \\
We here give some details of the derivation of (\ref{S, x}) and
(\ref{S, Lambda}).

The inverse vielbeins of the Poincar\'{e} group have the explicit integral
representations:
\begin{equation}
u_{\sigma \rho }\hspace{0.1 mm}^{\lambda \tau
}=\int_{0}^{1}dt(e^{t\phi })_{\sigma \rho }\hspace{0.1 mm}^{\lambda \tau },
\end{equation}
\begin{equation}
u_{\sigma }\hspace{0.1 mm}^{\lambda
}=\int_{0}^{1}dt(e^{t\phi })_{\sigma }\hspace{0.1 mm}^{\lambda },
\end{equation}
and
\begin{eqnarray}
u_{\mu \nu}\hspace{0.1 mm}^{\lambda}=\frac{1}{2}\int_{0}^{1}dt\int_{0}^{t}
du(e^{u\phi })_{\sigma }\hspace{0.1 mm}^{\lambda }a^{\zeta }C^{\sigma }
\hspace{0.1 mm}_{\zeta ,\omega \phi
}(e^{(t-u)\phi })_{\mu \nu }\hspace{0.1 mm}^{\omega \phi }\nonumber \\
=\frac{1}{2}a^{\zeta }C^{\lambda }\hspace{0.1 mm}_{\sigma ,\xi \eta }
\int_{0}^{1}dt\int_{0}^{t}du(e^{u\phi })_{\zeta}\hspace{0.1 mm}^{\sigma }
(e^{t\phi })_{\mu \nu }\hspace{0.1 mm}^{\xi \eta } \label{vielP}
\end{eqnarray}
where $\phi_\lambda \hspace{0.1 mm}^\rho $ is given in (\ref{fi}) and
\begin{equation}
\phi _{\xi \eta }\hspace{0.1 mm}^{\lambda \rho }=\frac{1}{2}
\lambda^{\mu \nu }C^{\lambda \rho }\hspace{0.1 mm}_{\mu \nu , \xi \eta }.
\end{equation}

By considering
\begin{eqnarray}
\{\Pi_{\mu \nu}, x ^{\lambda  }\}=\{\Pi_{\mu \nu}, u_{\rho }\hspace{0.1
mm}^{\lambda  }\}
a^{\rho }
=\frac{1}{2}u_{\mu \nu }\hspace{0.1
mm}^{\sigma  \rho }\{S_{\sigma
\rho }, x _{\lambda  }\}+u_{\mu \nu }\hspace{0.1 mm}^{\lambda \rho }x
_{\rho }
+u_{\mu \nu
}\hspace{0.1  mm}^{\lambda },
\end{eqnarray}
where, using the Poisson bracket
\begin{equation}
\{\Pi_{\mu \nu}, \lambda ^{\rho \tau  }\}=-(\delta _{\mu }\hspace{0.1
mm}^{\rho }
\delta _{\nu }\hspace{0.1 mm}^{\tau }-
\delta _{\nu }\hspace{0.1 mm}^{\rho }\delta _{\mu }\hspace{0.1 mm}^{\tau })
\end{equation}
we obtain
\begin{eqnarray}
\{\Pi_{\mu \nu}, u_{\rho }\hspace{0.1 mm}^{\lambda  }\}a^{\rho }
=-\int _0^1tdt\int _0^1du(e^{ut\phi })_{\sigma }\hspace{0.1 mm}^{\lambda }
C^{\sigma }\hspace{0.1 mm}_{\mu \nu ,\xi }
(e^{(1-u)t\phi })_{\rho }\hspace{0.1 mm}^{\xi }a^{\rho }
\nonumber \\
=-\frac{1}{2}\int _0^1dt\int _0^t(e^{u\phi})_{\mu \nu }
\hspace{0.1 mm}^{\xi \eta }
C^{\lambda }\hspace{0.1 mm}_{\xi \eta ,\zeta }
(e^{t\phi})_{\rho }\hspace{0.1 mm}^{\zeta }a^{\rho }
=u_{\mu \nu }\hspace{0.1 mm}^{\lambda \rho }x _{\rho }
+u_{\mu \nu }\hspace{0.1 mm}^{\lambda } \label {Pi, u}
\end{eqnarray}
and thus (\ref {S, x}).

For a Lorentz transformation
$\Lambda ^{\tau }\hspace{0.1 mm}_{\sigma  }$ we find:
\begin{eqnarray}
\{\Pi _{\mu \nu}, \Lambda ^{\tau }\hspace{0.1 mm}_{\sigma  }\}=
\int _0^1dt
(e^{-(1-t)\phi })_\lambda \hspace{0.1 mm}^\tau
C^{\lambda }\hspace{0.1 mm}_{\mu \nu ,\rho }
(e^{-t\phi})_\sigma \hspace{0.1 mm}^{\rho }
=\frac{1}{2}\Lambda ^{\tau }\hspace{0.1 mm}_{\zeta  }
C^{\zeta }\hspace{0.1 mm}_{\xi \eta , \sigma }
u_{\mu \nu }\hspace{0.1 mm}^{\xi \eta } \label{Pi, Lambda}
\end{eqnarray}
that is compared with:
\begin{eqnarray}
\{\Pi _{\mu \nu}, \Lambda ^{\tau }\hspace{0.1 mm}_{\sigma  }\}=
\frac{1}{2}u_{\mu \nu }\hspace{0.1 mm}^{\xi \eta }
\{M _{\xi \eta }, \Lambda ^{\tau }\hspace{0.1 mm}_{\sigma  }\}.
\end{eqnarray}
We conclude:
\begin{eqnarray}
\{M_{\mu \nu}, \Lambda ^{\tau }\hspace{0.1 mm}_{\sigma  }\}
=\{S_{\mu \nu}, \Lambda ^{\tau }\hspace{0.1 mm}_{\sigma  }\}
=C^{\zeta }\hspace{0.1 mm}_{\mu \nu ,\sigma }\Lambda ^{\tau }\hspace{0.1
mm}_{\zeta  }.
\end{eqnarray}

In order to deduce  (\ref{vielP}), (\ref{Pi, u}) and (\ref{Pi, Lambda}) we
used the identity
\be
(e^{t\phi})_\delta \hspace{0.1 mm} ^\gamma C^\delta \hspace{0.1 mm}_{\alpha
\epsilon }(e^{-t\phi})_\beta \hspace{0.1 mm} ^\epsilon
=(e^{t\phi })_\alpha \hspace{0.1 mm}^\delta
C^\gamma \hspace{0.1 mm}_{\delta \beta }
\ee
with $\phi $ defined in (\ref{phi}). This identity is a direct consequence
of the Jacobi identity of the structure constants
$C^{\gamma }\hspace{0.1 mm}_{\alpha \beta }$.

\end{document}